\documentclass[aps,pra,reprint,amsmath,amssymb,onecolumn,longbibliography,notitlepage,nofootinbib]{revtex4-1}

\usepackage{mathtools}
\usepackage{braket}
\usepackage{siunitx}
\usepackage{listings}
\usepackage[hidelinks]{hyperref}
\usepackage{setspace}
\usepackage{booktabs}
\usepackage{xcolor}
\usepackage{bm}
\usepackage{lineno}
\definecolor{lightblue}{RGB}{65,105,225}
\definecolor{orange}{RGB}{255,140,0}
\usepackage[margin=0.7in]{geometry} 

\usepackage{xparse}
\makeatletter
\renewcommand\@makefntext[1]{\parindent 1em\noindent
    \hb@xt@1.8em{%
        \hss\@makefnmark
    }#1
}
\renewcommand\@makefnmark{}
\renewcommand\@makefntext{\def\baselinestretch{1}\@textsuperscript{\@thefnmark}~}

\makeatother

\begin{document}

\title{TRON: Trainable, Architecture-Reconfigurable Random Optical Neural Networks}

\author{\small  Ziao~Wang$^{1,\dagger}$, Fei~Xia$^{1,2,\dagger,*}$, Logan~G.~Wright$^{3,4,6}$, Tatsuhiro~Onodera$^{3,4,5}$, Martin~Stein$^{3,4,6}$, Jianqi~Hu$^{1}$, Peter~L.~McMahon$^{3}$, Sylvain~Gigan$^{1,*}$}

\affiliation{$^1$Laboratoire Kastler Brossel, ENS-Universite PSL, CNRS, Sorbonne Université, Collège de France, Paris, 75005, France}
\affiliation{$^2$Department of Electrical Engineering and Computer Science, University of California Irvine, Irvine, 92617, CA, USA}
\affiliation{$^3$School of Applied and Engineering Physics, Cornell University, Ithaca, 14853, NY, USA}
\affiliation{$^4$NTT Physics and Informatics Laboratories, NTT Research, Inc., Sunnyvale, 94085, CA, USA}
\affiliation{$^5$NTT Physics and Informatics Laboratories, NTT Research, Inc., Sunnyvale, 94085, CA, USA}
\affiliation{$^6$Department of Applied Physics, Yale University, New Haven, CT 06511, USA}

\thanks{Corresponding authors' e-mails: fei.xia@uci.edu; sylvain.gigan@lkb.ens.fr}
\thanks{$^\dagger$These authors contributed equally to this work}


\maketitle

\begin{center}
\begin{minipage}{0.85\textwidth}
\small
\section*{Abstract}
Deep learning has triggered explosive growth in the demand for specialized hardware processors, thus motivating the development of scalable and reconfigurable computing substrates. Optical processors offer a fundamentally different computing paradigm, combining massive parallelism and ultrahigh bandwidth with the potential for substantial energy savings. However, progress has been constrained by the absence of scalable and reconfigurable architectures that can implement a broad class of network architectures. Here, we introduce TRON, a scalable and trainable optoelectronic deep optical neural network that exploits a multi-scattering medium and a DMD as a learnable, high-dimensional dense optical matrix multiplier, processing with fixed and tunable optical operations. We perform in-situ optimization of the optical parameters involved in the scattering process, together with automated neural architecture search (NAS) and optimization directly on optics. The experimental results demonstrate that in-situ NAS is essential to discover architectures that adapt to both the task and hardware constraints, establishing a viable path towards large-scale optical processors for next-generation machine learning and data-intensive computing. 
\end{minipage}
\end{center}

\section{Introduction}\label{sec1}
Deep learning \cite{lecun2015deep, schmidhuber2015deep, goodfellow2016deep} has become a central driver of progress in artificial intelligence, enabling breakthroughs that include, but are not limited to, computer vision \cite{he2016deep, krizhevsky2012imagenet, dosovitskiy2020image}, natural language processing \cite{liu2024deepseek, vaswani2017attention, devlin2019bert, brown2020language}, and scientific data analysis \cite{jumper2021highly, esteva2017dermatologist}. This rapid expansion has led to an unprecedented demand for specialized processing hardware \cite{jouppi2017datacenter, raina2009large, silvano2025survey}, accompanied by a sharp increase in computational costs and energy consumption \cite{AIcompute, strubell2019energy}. As deep neural networks continue to scale in size and complexity, improving the efficiency of the underlying computing platforms has emerged as a critical challenge \cite{kaplan2020scaling, patterson2021carbon}.

Optical computing offers a fundamentally different computing paradigm by performing analog computations directly in the optical domain. Owing to the intrinsic properties of light, optical systems naturally provide ultrahigh bandwidth, massive parallelism, and low latency, making them attractive candidates for energy-efficient information processing \cite{mcmahon2023physics, wetzstein2020inference, fu2024optical, brunner2025roadmap, shastri2021photonics,bente2025potential}. Inspired by the success of neural networks on electronic computing platforms, optical neural networks (ONNs) have been proposed as a class of optical processors capable of executing machine learning workloads in optical hardware \cite{mcmahon2023physics, wetzstein2020inference, fu2024optical, brunner2025roadmap, shastri2021photonics}. In principle, ONNs can combine the expressive power of neural networks with the physical advantages of optics to enable scalable and efficient computations. 

Recent experimental advances have demonstrated ONNs on integrated photonic and free-space platforms \cite{liu2021research, bernstein2023single, huang2021silicon, mengu2022all, li2022analysis, wang2022optical, lin2018all, chen2025optical,bente2025potential,kalinin2025analog,xia2024nonlinear,han2025optical,yildirim2024nonlinear,wanjura2025fully,wu2025field}. Integrated photonic implementations provide compact and stable optical processors for small-scale but fast matrix multiplications and have been applied to tasks such as optimization \cite{al2025programmable}, advanced machine learning \cite{wanjura2025fully,wu2025field,hua2025integrated,ahmed2025universal}, optical communications, and signal processing \cite{huang2021silicon}, while free-space optical platforms offer high-dimensional processing for imaging \cite{mengu2022all, li2022analysis}, sensing \cite{wang2022optical, xia2024nonlinear}, and standard machine learning tasks \cite{lin2018all, chen2025optical, kalinin2025analog, yildirim2024nonlinear}. These demonstrations underscore the potential of optical processors to perform neural network computations by encoding information on optical signals. Despite this progress, most of the ONNs demonstrated to date remain limited in scalability and flexibility. Many optical processors are based on fixed or relatively shallow network architectures, usually with fewer than three layers, and include only a small number of input neurons (often under a few hundred). Diffractive neural networks made using static optical elements, for example, are optimized for one type of task and lack reconfigurability after fabrication \cite{wetzstein2020inference, lin2018all, shastri2021photonics}. Programmable ONNs \cite{bernstein2023single,wang2022optical, choi2025transferable} expand the diversity of computing tasks but are still typically limited to a few layers, making their applicability to large-scale deep learning problems further limited \cite{shen2017deep, hughes2018training}. Scaling up matrix-vector multiplication, which is the core operation in deep neural networks, remains also a major challenge for optical processors, particularly in closed-loop training and high-speed operation due to hardware speed limitations\cite{wetzstein2020inference, shastri2021photonics, momeni2025training}. All these limitations hinder the ability of current ONNs to represent complex functions and to generalize in various tasks that require deeper and wider neural networks\cite{wetzstein2020inference, shastri2021photonics, bahri2024explaining}. 

Here, we present an alternative approach that overcomes the above limitations by exploiting the high-dimensional, analog, and complex weight mixing intrinsic to optical scattering media. Complex scattering media, characterized by strong spatial heterogeneity in their optical properties, have previously been used to realize large-scale optical random projections \cite{gigan2022imaging} for kernel machines \cite{saade2016random}, linear operators \cite{matthes2019optical, ohana2020kernel} and learning algorithms based on random features  \cite{launay2020direct, wang2024streamlined}. However, such systems have largely been limited to fixed, untrainable transformations. In this work, we transform fixed optical scattering into a programmable and trainable computational resource by actively controlling the high-dimensional, input-dependent mixing of light through a complex scattering medium using a spatial light modulator. This enables reconfigurable, high-dimensional matrix-vector multiplication, while intensity detection at the camera introduces an inherent nonlinearity. We refer to this architecture as a trainable random optical neural network (TRON), as it merges the trainability of a neural network with the properties of an optical random scattering medium. By treating the scattering-based system as a flexible computational layer, we demonstrate in situ reconfiguration of both network depth and width, enabling automated optimization of network architecture and weights directly on the physical system. Our approach combines massive optical parallelism with scalability, reconfigurability, and efficient trainability via backpropagation, which offloads the dominant forward passes workload to the optical hardware, while retaining electronic control and training-time gradient computation. Using this framework, we demonstrate accurate in situ inference on both scientific and real-world datasets, establishing a pathway towards scalable optical processors for large-scale machine learning and other data-intensive computing tasks.

\section{Results}\label{sec2}
\subsection{Scalable, automated and hardware-efficient architecture design for deep TRON tailored for computing tasks}\label{sec3}
\begin{figure*}[!htbp]
  \centering
  {  \makebox[\textwidth]{\resizebox{1\linewidth}{!}{\includegraphics{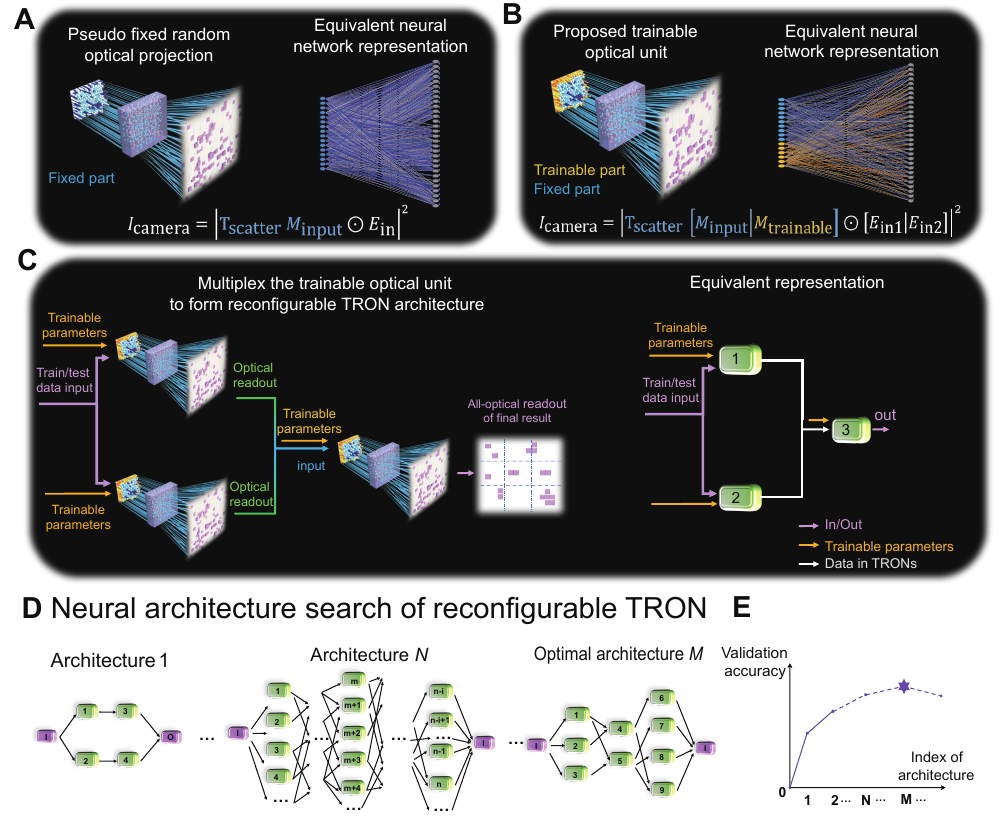}}}}

  \caption{
 Concept of a trainable, reconfigurable TRON. \textbf{A.} Conventional large-scale random optical projection using a complex scattering medium. An input field is modulated, propagated through the medium, and its intensity is recorded. This fixed linear mapping corresponds to a single neural network layer with random, non-trainable weights. \textbf{B.} Proposed trainable optical unit. A subset of the modulation degrees of freedom (for instance, specific pixels on the DMD) is designated as learnable parameters, enabling the complex medium to implement a tunable transformation and thus serve as an TRON building block with trainable weights. \textbf{C.} By time-multiplexing this trainable unit and feeding the optical outputs back as inputs at subsequent time steps, our platform realizes a reconfigurable TRON with a flexible architecture. \textbf{D.} The TRON architecture is optimized \textit{in situ} via NAS executed directly on the physical hardware for a target task. \textbf{E.} This in-situ NAS yields an optimal TRON architecture based on validation accuracy.
  }
  \label{fig:1}
\end{figure*}

To introduce our concept, we start from a conventional optical computing configuration based on a randomly scattering medium, in which the input data is encoded by a specialized SLM that has binary pixels as input named the digital micromirror device (DMD), then mapped through the fixed scattering medium and detected in intensity, implementing a large-scale single-layer random projection with no trainable parameters (Fig.~\ref{fig:1}A). Here, instead of using the randomly scattering medium only as a fixed random projector, we exploit the inherent analog, complex-valued ``optical weights" with up to 10$^{12}$ in dimension (Supplementary Section~1) to design an TRON with reconfigurable randomized matrix-vector multiplication. The general principle starts by partitioning the modulator pixels into two sections: one for trainable parameters optimized for specific tasks, and the other as input for training or inference data. This partitioning allows trainable complex-weight mixing in the medium, followed by intensity detection as a nonlinear operation, effectively implementing a trainable one-layer TRON (Fig.~\ref{fig:1}B, Supplementary Section~4). 

Although such optical random matrix-vector multiplication offers potential scalability due to its simple design, a single trainable TRON module remains limited in expressiveness. One pass through the module provides only a shallow model of the input-output mapping. To further improve the expressiveness of TRON and more accurately approximate task-specific functions, one of the most effective strategies is to reconfigure dynamically the network architecture. This enables optimization beyond the current weight and bias limitations imposed by the materials and system, as evidenced by the progress in various state-of-the-art deep learning architectures that achieve superior computational performance across multiple tasks \cite{he2016deep, huang2017densely, szegedy2017inception}. We propose to implement this optically using the same trainable optical module, i.e., the same system. Instead of relying on the spatial multiplexing of multiple optical systems, which is straightforward to realize but costly, we adopt time multiplexing in the trainable TRON, which means that the same system is reused at different times. This enables flexible scaling in both depth and width within a single optical setup (Fig.~\ref{fig:1}C). It is also inspired by optical reservoir computing, where a single optical system is used to implement a recurrent neural network \cite{rafayelyan2020large, dong2019optical, van2017advances}. However, in our case, we enable far greater trainability in the system, so its design flexibility and computational capabilities go well beyond conventional physical reservoir computing and become closer to a conventional deep neural network. By regulating the data flow into the same optical system at different time steps, we construct a TRON architecture whose depth, width, and connectivity can all vary. For example, to increase the depth, we sequentially use the output image from one camera capture as the input for subsequent stages on the DMD, whereas for wider configurations, we present the same input image to the DMD at different times. 

Previous ONN implementations have predominantly depended on hand-designed architectures, which either demanded substantial trial-and-error or used designs optimized solely in numerical simulations before being transferred to hardware. This often leads to suboptimal performance because of the simulation-reality gap. To overcome both limitations and to better exploit the reconfigurability of our platform, we perform an optical neural architecture search (NAS)~\cite{zoph2016neural, elsken2019neural} on the TRON hardware in situ (Fig.~\ref{fig:1}D), thereby automating the architecture design directly on the physical system. In our optical NAS approach for TRONs, we define a search space of architectural components that are natural to the optical setup, including the number of time-multiplexed passes, their connections, and the associated data-parameter pixel partition. Candidate architectures within this search space are implemented on the optical hardware and evaluated based on their performance on a validation set (split from the training set). In practice, we frequently find that the architecture chosen by highest validation accuracy, defined as the optimal architecture, is not the one with the greatest depth or width, as shown in Fig.~\ref{fig:1}E. The resulting optimal architecture is subsequently assessed on the held-out test set, which remains untouched throughout the in-situ NAS process. In this regard, our method is consistent with the general NAS framework in machine learning~\cite{elsken2019neural} and with hardware-aware NAS for conventional accelerators~\cite{chitty2022neural}, but it is deployed in an optical domain where the ``optimal" architecture can only be robustly identified by interrogating the optical device itself. Using in-situ NAS to design task-specific TRONs enables a more effective utilization of the optical hardware’s capabilities. This automated optical machine-learning  approach not only enhances the performance of our TRON, as detailed in later sections, but also uncovers new architectures that had previously been difficult to obtain through manual design.

\subsection{Hybrid training for automated optical machine learning for in-situ inference}\label{subsec2}
\begin{figure*}
  \centering
  {  \makebox[\textwidth]{\resizebox{1\linewidth}{!}{\includegraphics{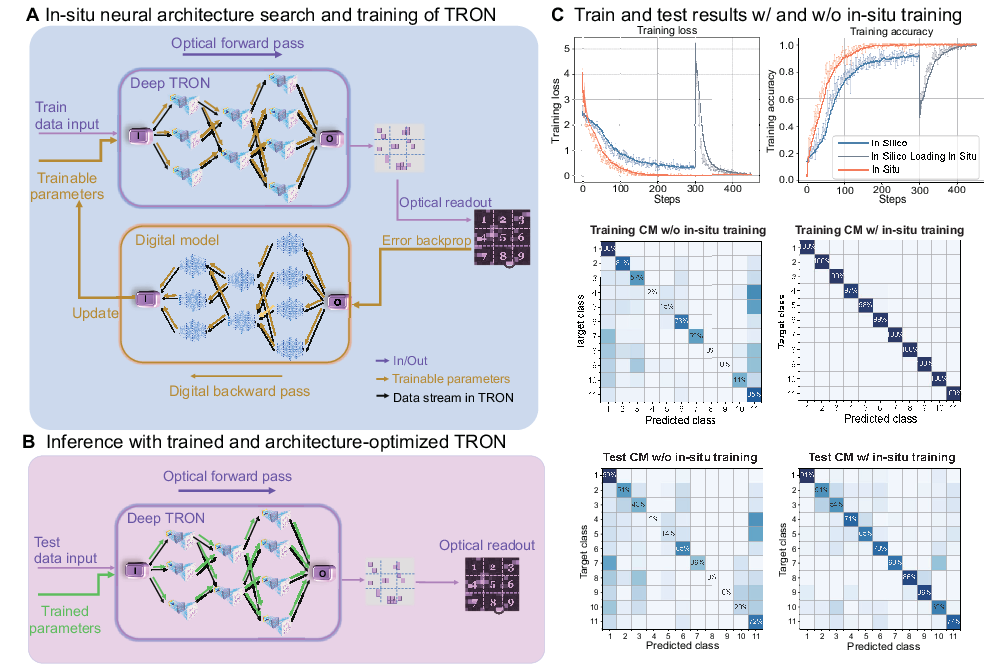}}}}
  \caption{
  Hybrid training enables optical inference and reveals the necessity of in-situ training. \textbf{A.} Hybrid training loop for TRON optimization. Each iteration combines one optical forward pass to produce an optical readout, and one digital backward pass, computed through Physics-Aware Training (PAT), enabling gradient flow and updates of the trainable parameters. \textbf{B.} In-situ inference with the trained TRON, where only the optical forward pass is executed using the trained parameters. The final predictions is obtained by optical readout without any digital layers, therefore isolating inference performance to the TRON itself. \textbf{C.} Representative training dynamics and confusion matrices (CMs) on the 3D MedMNIST (organ) under two different training protocols: in silico training followed by parameter loading and in-situ training, and fully in-situ training. Although fully in-situ training converges in fewer steps, the in-silico-then-in-situ strategy can reduce overall wall-clock time when the task only uses $2\%$ the optical maximal throughput. A significant degradation is observed immediately after deploying the in-silico-trained model, and is rapidly recovered and further improved by in-situ training, highlighting the in-situ training is essential to achieve optimal performance.
  }
  \label{fig:2}
\end{figure*}

To efficiently optimize the reconfigurable TRON architecture proposed as an NAS candidate, we adopt a hybrid training approach. As shown in Fig.~\ref{fig:2}A, our training procedure comprises an optical forward pass and a digital backward pass, following the paradigm of physics-aware training~\cite{wright2022deep}. During the forward pass, the data propagate multiple times through the optical setup to produce the optical readout, with each TRON node representing one single pass through the setup. During the backward pass, we employ a digital model to approximate the gradients and adjust the trainable parameters on the DMD. For the digital model, we utilize prior knowledge of the physical model in the form of a transmission matrix \cite{popoff2010measuring} for the input–output mapping, expressed as \( I_{\text{camera}} = |R I_{\text{DMD}}|^2 + I_{\text{noise}} \). Here, \( R \) denotes the transmission matrix mapping from the DMD patterns \( I_{\text{DMD}} \) to the camera patterns \( I_{\text{camera}} \), while \( I_{\text{noise}} \) represents the system noise. This optics-aware formulation of the digital model enables us to compute gradients that are consistent with the physical forward process while still allowing error backpropagation to be carried out in practice. Importantly, although hybrid training introduces an extra computational cost for evaluating digital gradients, this overhead is limited to the training stage. After the TRON has converged, inference relies solely on the optical forward propagation (Fig.~\ref{fig:2}B), and the final prediction is read directly from the camera output. No further digital layers are required.

We evaluate the proposed TRON training framework, as well as the importance of in-situ training, on the 3D Organ Classification task (Fig.~\ref{fig:2}C) using the MedMNIST dataset~\cite{yang2023medmnist}. This dataset features an input neuron count (at a scale of 10$^4$) that is an order of magnitude larger than that of most state-of-the-art optical neural networks. We consider two training protocols: (1) fully in-silico training, and (2) in-silico pretraining followed by parameter loading and subsequent in-situ training. When a model trained purely in silico is directly deployed onto the optical system without any in situ training, we observe a significant simulation-reality gap. After deployment, the loss jumps to levels similar to or even worse than those observed with random initialization, and the accuracy likewise declines. This deterioration is also visible in the confusion matrices in Fig.~\ref{fig:2}C. By contrast, if training is resumed in situ after loading, performance quickly rebounds and then improves further. Additional transfer-learning experiments are reported in Supplementary Section~5. Collectively, these findings show that in situ training on the physical hardware is effectively required to counteract system imperfections, environmental noise, and mismatches between the model and the hardware system.

We observe that although the fully in situ protocol converges in fewer optimization steps than the hybrid ``in silico + in situ" protocol, we opt for the hybrid approach in the remaining experiments because of its superior overall wall-clock efficiency. The horizontal axis in Fig.~\ref{fig:2}C counts optimization steps rather than elapsed time, and for tasks such as MedMINST-3D-Organ, the problem dimensionality uses only about 2\% of the maximum data throughput of the optical platform. In this low-utilization regime, the per-step latency of the optical loop (primarily determined by optical/electronic I/O) can actually surpass that of the GPU-based digital model. Consequently, using in silico training as an initial warm-up phase, followed by in situ training, strikes a practical balance between training speed and final model quality, while still leveraging the distinctive benefits of the TRON hardware during in situ NAS.

\subsection{Application of deep TRON with inference with all-optical readout}\label{subsec3}
To demonstrate the regime in which TRON is scientifically consequential, we deliberately target tasks whose input dimensionality far exceeds that of typical ONN benchmarks. We therefore select two high-dimensional real-world problems: volumetric 3D CT classification and genomics classification from RNA expression profiles. These tasks move beyond the low-dimensional image benchmarks which widely examined by optical computing, and directly probe the regime matched to scattering-based optics, i.e., large-scale projections executed at high throughput.

\subsubsection{Large-scale 3D medical CT image classification}\label{subsubsec3-1}
The domain of 3D image processing, especially in clinical settings such as CT imaging, poses distinct challenges arising from the complexity and variability of anatomical structures. Reliable classification of these images is essential for a range of medical tasks, including disease diagnosis, planning therapeutic interventions, and monitoring patients over time. Here, we demonstrate the ability to process a real-world 3D medical CT dataset with image dimensions of $2.2\times 10^4$, which demands substantially more computational resources than the widely used datasets MNIST or FashionMNIST. This unusually high input dimensionality directly matches the regime targeted by TRON, while the largely binary contrast structure of CT data also fits naturally with our optical interface. The larger dataset scale therefore underscores both the medical relevance of the task and the scalability and flexibility of our proposed deep TRON architecture for handling complex medical imaging data. The larger dataset scale underscores the scalability and flexibility of our proposed deep TRON architecture for handling complex medical imaging data.

By leveraging in situ training and in situ NAS, we identify an optimal deep TRON architecture, which can then be implemented on a single optical unit and subsequently trained. This strategy not only automates the architecture design process but also guarantees that the resulting network is well adapted to the 3D medical imaging dataset. Figure~\ref{fig:3}A illustrates the inference pipeline, in which the CT input and trained parameters configure the TRON for optical forward propagation, yielding the optical readout. Class scores are subsequently obtained from speckle intensity statistics defined for specific regions, wherein the locations are fixed in advance and the associated parameters are optimized to enable the optical readout. To evaluate the benefits of the discovered architecture, we compare four TRON configurations that share the same total resource budget of 15 units, i.e., 15 TRON nodes (Fig.~\ref{fig:3}B). These configurations consist of a deep network (in series), a wide network (in parallel), a standard NN (5 units per layer, 3 layers), and the  architecture obtained via our NAS. The confusion matrices and corresponding ROC curves indicate that the architecture selected via in-situ NAS outperforms the other candidates. Furthermore, we conducted hyperparameter optimization using validation sets to further enhance model performance.

\begin{figure*}[!htbp]
  \centering
  {\includegraphics[width=0.9\linewidth]{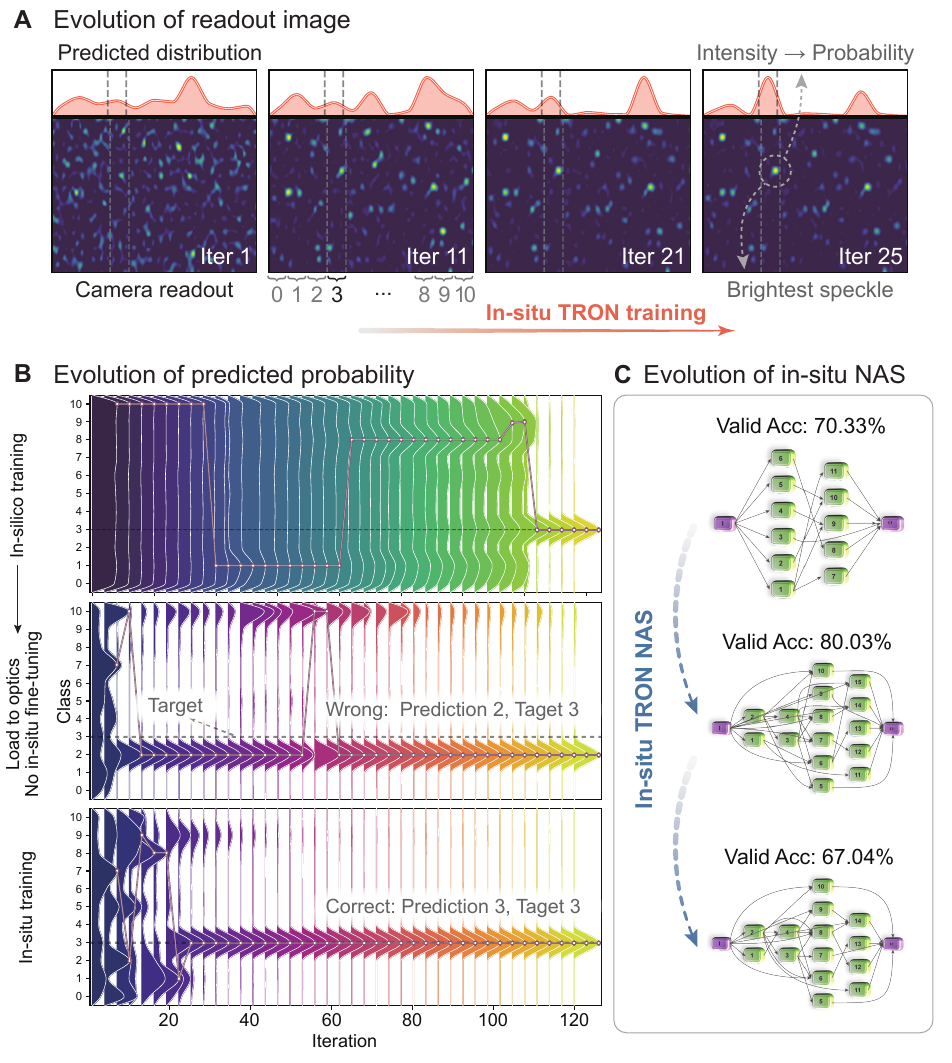}}
  \caption{ 
  Evolution of the in situ training and in situ NAS on the 3D CT task. \textbf{A.} Evolution of the optical readout during in-situ TRON training. Each snapshot shows the optical readout at a given iteration, where the region-wise speckle intensities present the predicted probability. As the in situ training proceeds, the peak speckle in the target-class region becomes brighter and finally the dominant feature. \textbf{B.} Probability evolution across iterations highlights the deployment mismatch and the role of in-situ training. Each vertical profile corresponds to the predicted class-probability distribution at one iteration (dashed line: target class). For each in silico iteration (top), the corresponding parameters are loaded onto the optics and evaluated by the optics without in-situ training (middle), which can yield an incorrect prediction after deployment. The predictions during in-situ training (bottom) predicts correct classification and converge faster. \textbf{C.} Representative evolution of in-situ TRON NAS. The search starts from relatively simpler candidates, explores more complex connectivity, and then refines around high-performing candidates. Reported accuracies in this figure are validation accuracies, as the NAS procedure uses the validation set for selection while keeping the test set untouched.
  }
  \label{fig:4} 
\end{figure*}

Next, we investigate how in-situ training shapes the all-optical readout and why it is essential for robust deployment in this 3D CT scan task. Figure~\ref{fig:4}A shows that in-situ training progressively reallocates optical energy towards the target class: as the iterations proceed, the peak intensity within the target-class region increases relative to other regions and eventually dominates, in line with rising confidence under the readout rule (see Supplementary Section~7 for details). Figure~\ref{fig:4}B offers a complementary representation in the probability space. While in-silico training can achieve correct predictions, directly transferring those parameters to the optical hardware without in-situ fine-tuning can lead to misclassification, revealing a deployment mismatch consistent with Fig.~\ref{fig:2}C. In contrast, in-situ training yields correct classifications and converges more rapidly, demonstrating that on-hardware training is practically necessary for high-fidelity optical inference on this task. Lastly, Fig.~\ref{fig:4}C outlines the trajectory of the in-situ TRON NAS (details in Supplementary Section~6). The search initially explores richer connectivity patterns and then narrows down to high-performing candidates. We emphasize that model selection is conducted on the validation set (Valid Acc.), with the test set kept untouched throughout in-situ NAS. Given together, these results show that TRON performs strongly because the architecture, trainable optical degrees of freedom, and readout formation are all optimized under real device constraints rather than idealized numerical proxies.

\begin{figure*}[!htbp]
  \centering
  {\includegraphics[width=0.99\linewidth]{figures/MainFigs/Figure3_organ_v5.pdf}}
  \caption{
  Large-scale 3D medical CT classification with a deep TRON and all-optical readout. \textbf{A.} Task overview and inference dataflow following the training paradigm of Fig.~\ref{fig:2}A. The optical forward pass produces a camera optical readout. The readout is partitioned into $N$ class regions and the region-wise intensity yields the predicted probability. \textbf{B.} Performance comparison among different architectures with a fixed budget of 15 TRON units. A deep architecture (1 unit per layer, 15 layers), a wide architecture (15 units per layer, 1 layer), a conventional NN architecture (5 units per layer, 3 layers), and the architecture selected by in-situ TRON NAS. For each architecture (columns), we show the architecture (top), the held-out test confusion matrix (middle), and the corresponding multi-class ROC curves (bottom). Under identical unit budget, the NAS-selected architecture achieves the highest test performance among the compared designs.
  }
  \label{fig:3}
\end{figure*}

Table~\ref{tab:comparison} presents a summary of these results in comparison with other digital baselines that were trained and evaluated on the same dataset. It should be emphasized that the classification of medical 3D CT images is widely recognized as a difficult problem. Nevertheless, with a binarized optical input, our deep TRON reaches $79.02\%$ test accuracy on the 3D organ task, which is comparable to ResNet-18 + 2.5D ($78.80\%$) and ResNet-50 + 2.5D ($76.90\%$). A key feature of deep TRON is that the dominant share of its forward-pass workload is executed by optical propagation. Although these operation counts are not directly comparable to programmable FLOPs, because the dominant optical transformation is realized by a fixed transmission matrix and the optical input and readout are quantized, they nevertheless quantify the scale of information processing physically carried by the optical substrate. For the 3D MedMNIST task, this corresponds to approximately $8 \times 10^9$ fixed optical operations and $3\times 10^5$ tunable optical operations per sample, placing TRON in the same operation-count regime as standard deep digital baselines. The end-to-end throughput additionally depends on optoelectronic I/O, time-multiplexing, and task-specific utilization (Methods F). Details about the throughput of our system and a contextual comparison to an on-chip photonic processor under shared metrics are provided in Supplementary Section~4. Additionally, a detailed analysis of trainable parameters and computing operations across the digital and optical domains highlights the potential optical advantage in energy consumption (Supplementary Section~10) and speed as the system scales. With up to $99.99\%$ of the forward-pass workload carried out optically, these results support the promise of energy-efficient and scalable optical processing.

\begin{table}[ht]
  \centering

  \label{tab:comparison}
  \begingroup
  \setlength{\tabcolsep}{6pt}   
  \renewcommand{\arraystretch}{1.2} 
  \begin{tabular*}{\textwidth}{@{\extracolsep{\fill}} 
      c 
      c 
      c 
      p{0.22\textwidth} 
      p{0.22\textwidth} 
    @{}}
    \hline
    \textbf{Method} 
      & \textbf{3D Organ} 
      & \textbf{3D Fracture} 
      &
       \begin{minipage}[t]{0.22\textwidth}
      \textbf{\# Params} 
       \end{minipage} 
      & 
      \begin{minipage}[t]{0.16\textwidth}
        \textbf{\# Ops/sample}
      \end{minipage} \\
    \hline
    TRON (ours) 
      & 0.7902
      & 0.5167 
      & 
      \begin{minipage}[t]{0.22\textwidth}
        $\sim 2\times 10^{5}$\\
        (\textcolor{purple}{$2.8\times10^{5}$}\\
        \textcolor{orange}{$+\;0.7\times10^{5}$}\\
        \textcolor{lightblue}{$+\;5.3\times10^{8}$)}
      \end{minipage}
      & 
      \begin{minipage}[t]{0.16\textwidth}
        \textcolor{orange}{$\sim 8\times10^{9}$}\\
        \textcolor{purple}{$+\;3\times10^{5}$}
      \end{minipage} \\
    \hline
    1‐hidden layer MLP (ReLU) 
      & 0.1100 
      & 0.4500 
      & 
     \begin{minipage}[t]{0.22\textwidth}
       $4.39\times10^{4}$ 
       \end{minipage} 
      &
      \begin{minipage}[t]{0.16\textwidth}
       $4.39\times10^{4}$ 
       \end{minipage} \\
  
    ResNet‐50 + 2.5D 
      & 0.7690 
      & 0.3970 
      &  
           \begin{minipage}[t]{0.22\textwidth}
       $4.58\times10^{7}$ 
       \end{minipage} 
      &      
      \begin{minipage}[t]{0.16\textwidth}
       $3.8\times10^{9}$
       \end{minipage} \\
  
    ResNet‐18 + 2.5D 
      & 0.7880 
      & 0.4510 
      & 
       \begin{minipage}[t]{0.22\textwidth}
       $3.17\times10^{7}$ 
       \end{minipage} 
      & 
       \begin{minipage}[t]{0.16\textwidth}
     $1.9\times10^{9}$
       \end{minipage} \\
    
    ResNet‐50 + 3D 
      & 0.8830 
      & 0.4940 
      & 
       \begin{minipage}[t]{0.22\textwidth}
        $4.64\times10^{7}$
       \end{minipage} 
      &   
      \begin{minipage}[t]{0.16\textwidth}
      $3.8\times10^{9}$ 
       \end{minipage}\\
    
    ResNet‐18 + 3D 
      & 0.9070 
      & 0.5080 
      & 
        \begin{minipage}[t]{0.22\textwidth}
        $3.33\times10^{7}$ 
       \end{minipage} 
      &   
      \begin{minipage}[t]{0.16\textwidth}
     $1.8\times10^{9}$
       \end{minipage} \\
    \hline
  \end{tabular*}
  \endgroup
    \caption{Comparison with other state-of-the-art digital neural network architectures trained for 3D MedMNIST task. Params: parameters, Ops: forward-pass operations per sample under the convention defined in Method F. \textcolor{purple}{Text in purple} means parameters or operations in the \textcolor{purple}{digital quantities}; \textcolor{orange}{orange} means parameters or operations in the \textcolor{orange}{optical domain and are tunable};    \textcolor{lightblue}{text in blue} means parameters or operations in the \textcolor{lightblue}{optical domain and are fixed}. These operation counts exclude electronic I/O, device control.
    }
\end{table}

\subsubsection{Computing with RNA sequence for cell type-specific disease classification}\label{subsubsec3-2}
In the second demonstration, we move beyond image processing and investigate sequential data processing using light. We selected a dataset from genomics, a domain that requires powerful and precise methods to analyze and categorize high-dimensional, complex data. Gene sequence classification is particularly important because it can reveal patterns of gene expression, illuminate cellular mechanisms, and provide insights into the origins and progression of diseases, such as cancer. To illustrate the flexibility and potential of our TRON for addressing complex, non-image-based, real-world problems, we extended our study to include the classification of gene sequence data.

\begin{figure*}[!htbp]
  \centering
  {\includegraphics[width=0.9\linewidth]{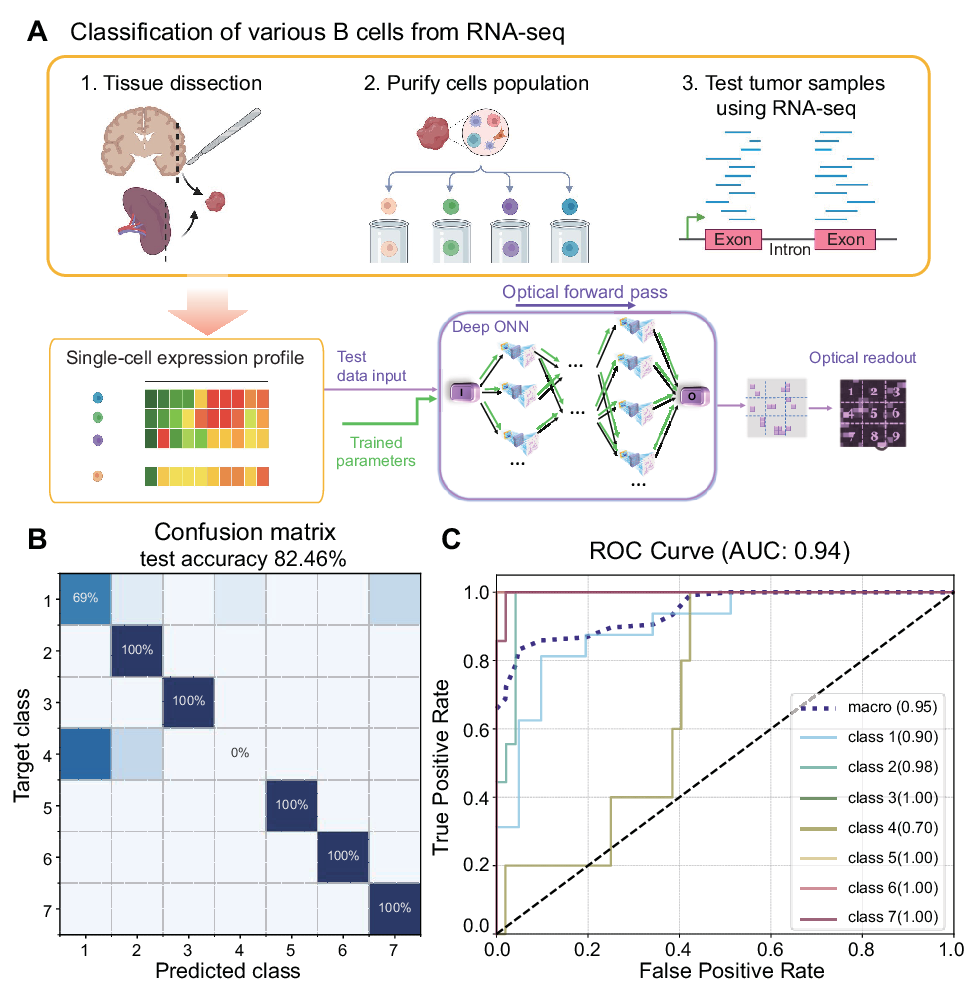}}
  \caption{Demonstration of the architecture-optimized TRON for gene classification. (a) Conceptual illustration of the RNAseq
dataset, comprising diverse B cells extracted from diseased tissues. (b) Confusion matrix for the architecture optimized
TRON tailored to this task. (c) Receiver operating characteristic curve (ROC) result with Area Under the Curve (AUC) the architecture optimized
TRON tailored to this task.}
  \label{fig:5}
\end{figure*}

The dataset used in this study originated from the Curated Microarray Database (CuMiDa)\cite{feltes2019cumida}, a comprehensive resource containing 78 carefully selected cancer microarray datasets. CuMiDa was derived from more than 30,000 studies available in the Gene Expression Omnibus (GEO) database and was specifically assembled to support machine learning research in this field (Fig. \ref{fig:5}a).

Our study utilized a leukemia dataset from the CuMiDa repository, which includes seven distinct classes and comprises 222,284 genes. This dataset enabled us to evaluate TRON’s capability to process high-dimensional, non-image data while simultaneously tackling a key problem in bioinformatics and genomics. We obtained a test accuracy of 82.46$\%$ and an AUC of 0.94, which are on par with those of widely used machine learning architectures (Supplementary Fig. S10), while keeping most of the computations in the optical domain.

By applying our TRON to this complex task, we sought to highlight its flexibility and effectiveness in data processing scenarios, specifically, 3D image analysis and genome analysis, which extend beyond the 2D image processing that has been the primary focus of prior optical computing studies. Our method demonstrated the TRON’s ability to manage high-dimensional data, a critical advantage in biomedical contexts, where data are both vast and intricate. This work further emphasizes the promise of hybrid opto-electronic neural networks for addressing a broad spectrum of applications, particularly in the rapidly growing fields of scientific research and real-world applications.

\section{Discussion}\label{sec12}
A central outcome of this work is the demonstration of in situ NAS performed directly on a reconfigurable TRON, whereby architectures are optimized and selected under the optical hardware response rather than in an idealized numerical model; we close the loop with the physical hardware response, allowing the search to account for non-idealities (e.g., noise, drift, and dynamic range) that strongly modulate the effective capacity of a given architecture. This approach enables the automated discovery of task-specific TRON configurations.


Notably, the architectures identified in situ in our case consistently favor skip connections over purely deep serial chains (Supplementary Fig.~S9). We interpret this as a direct signature of algorithm-hardware synergy. Because each unit reencodes its signals through a 1-bit DMD, a naive serial depth accumulates precision loss across layers. The shortcut in the architecture compensates for this cumulative quantization effect. In TRON, in-situ NAS not only tunes the architecture in the abstract but also discovers dataflows matched to the quantized optical hardware, compensating for physical limitations that simulation-based design sometimes overlooks.  

This mechanism clarifies why TRON performs strongly on both tasks, despite the simplicity of the optical part. Multiple scattering supplies an extremely high-dimensional feature map, and the trainable optical freedoms turns this map into a task-adaptable transformation. In situ optimization aligns the controllable degrees of freedom and readout statistics to the task under the true hardware response, absorbing mismatch, noise, and readout non-idealities directly into the learned solution. The resulting performance is thus not a simple consequence of random media, but of coupling large-scale optical mixing with trainable parameters, hardware-aware optimization and hardware-aware architectures. This combination defines a concrete design principle for future ONN that strong optical hardware becomes even better when its architecture and parameters are optimized together under the physical constraints of the device.

Several challenges lie ahead. The overall search and training throughput/scalability is currently bounded by optoelectronic bandwidth and by the calibration/stability requirements of repeated physical measurements. Advancements in optical device innovation will play an instrumental role in enabling rapid optoelectronic detection and optoelectronic conversion, such as metasurfaces for tunable scattering matrices \cite{brongersma2025second}, high-speed light modulators \cite{brongersma2025second,rocha2024fast} and high-speed image sensors \cite{gao2014single, nakagawa2014sequentially}. As these technologies progress, we can anticipate improvements in the speed, tunability, and efficiency of TRONs, ultimately leading to enhanced scalability of our design. This progress will contribute to the development of broader computing platforms, where TRONs can be integrated with conventional electronic systems. As the field of unconventional computing platforms evolves, the development of deep and reconfigurable TRONs holds the potential to inspire new algorithms and problem-solving techniques specifically designed for these platforms. These developments have the potential to substantially transform the future landscape of unconventional computing, a field whose architectures have long suffered from limited tunability.

\section{Methods}\label{sec11}

\subsection{Optical setup and experimental description} \label{opu-conf}

Our system leverages light scattering as the core computational mechanism and builds upon our previous optical processing unit (OPU) design \cite{saade2016random}. It serves as an elementary computing block and is iteratively used in our TRON. Primarily, the setup comprises a continuous-wave laser, a DMD for encoding the information, a disordered scattering medium, and a CMOS imaging camera.

A diode-pumped solid-state laser source centered around the $532.3$ nm wavelength delivers light through a fiber to a telescope. The telescope expands the laser beam to illuminate a DMD featuring a $912\times 1140$ micro-mirror array arranged in a diamond pixel configuration. Each micro-mirror pixel in the array encodes 1-bit spatial information into the light beam by adjusting its angular position ($-12^{\circ}$ or $12^{\circ}$), determining whether the light at this pixel reflects towards the scattering medium. After spatially encoded light passes through the scattering medium, the speckle intensity distribution is collected by a pipelined CMOS image sensor, with $2048\times 1088$ pixels and 8-bit quantization.

We encode any numerical vector with real values $\mathbf{x}$ from a dataset into the light beam using a 1-bit precision DMD. As detailed in the Supplementary~Alg. 2, each dimension of the vector normalized to the range $[0, 1]$ is binarized into either $1$ (towards the scattering medium) or $0$ (not towards) through a custom threshold $t_i$, which is initially set at $0.5$. The threshold is specifically adjusted for each dimension during the training process by treating it as a parameter. The resulting vector $\tilde{\mathbf{x}}$ is then reshaped into a 2D array of $[w, h]$, mirroring the original 2D data structure for image tasks. If the data are not originally 2D, we reshape them to facilitate display on the DMD while preserving the internal structure of the data to the greatest extent possible. The binarized optical parameters $\mathrm{Sign}(\bm{\theta})$ are then arranged around the border of the 2D array, forming $[\tilde{\mathbf{x}}_i; \mathrm{Sign}(\bm{\theta})]$. In practice, multiple pixels (macro-pixels) on the DMD are used to represent each input element, thereby minimizing crosstalk. The macro-pixel size is selected to ensure that the 2D array maximally occupies the DMD region when centered, while maintaining the aspect ratio. The remaining border pixels on the DMD are set to $0$ to prevent static bias in the output speckles. After sending the data through the scattering medium, we capture a snapshot of the speckle patterns within a predefined region of interest on the camera. To increase the operating speed of the camera, row-windowing in the height direction of the camera is utilized by programming the sensor registers. Each frame only reads out the centered $[2048, h]$ pixels, and the speckle image $\bm{o}$ in the middle of the shape $[w, h]$ is stored. Afterward, the speckle images of an 8-bit depth ($0$ to $255$) are rescaled to $0$ to $1$ again and are used as inputs for the subsequent computation of optical units.

The benefits of the setup come from the dimensionality of the data, the speed at which the random projections are made, and the low power consumption. The input $\tilde{\bm{x}}$ and output $\bm{o}$ scale to dimensions of 1 million and 2 million, respectively. In the usage of both full-size $\tilde{\bm{x}}$ and $\bm{o}$, the entire device can run at near $340$ Hz while the maximum frame rate can reach $1.8$ kHz. At full size, the DMD runs at $2880$ Hz for the continuous streaming mode and $4220$ Hz for the burst mode with maximal $48$ binary patterns at once, while the camera runs at $\sim340$ Hz. However, with the row-windowing mode, the frame rate of the camera can be increased up to $5.5$ kHz. We characterize the temporal stability of the optical setup over a duration of 3 h (Supplementary Section 2).

\subsection{In-situ NAS of TRON}
We adopt the concept of NAS in the design of our reconfigurable deep TRON architecture to automate the development of ONNs, with the aim of achieving performance that meets or exceeds hand-crafted architectures (Supplementary Alg.~3-4). We control, simulate, and train our system using Python. In our Python implementation, we utilize the Optuna package, which offers state-of-the-art algorithms for sampling hyperparameters and pruning unpromising trials during the development of neural network (NN) architectures. We employ the Tree-Structured Parzen Estimator (TPESampler), a Bayesian optimization-based algorithm that iteratively constructs a probabilistic model of the relationship between hyperparameter values and objective functions to facilitate global optimization of hyperparameters, including the learning rate, number of layers, and number of hidden units in each layer. In the pruning stage of the NAS, we select MedianPruner, a pruner that employs the median stopping rule when the trial's best intermediate result is inferior to the median of intermediate results from previous trials at the same step. Typical optimal architectures involve 3-5 TRON layers with a maximum of 15 TRON units in total, trained using the Adam optimizer, the cross-entropy loss function, and learning rates of  10$^{-3}$. For further details, refer to Supplementary Section~8.

For each NAS candidate, the performance of the TRON architecture is evaluated after one two-stage hybrid training process. For every trial, we first perform an in-silico warm-up training for 20 epochs using the digital twin, after which the obtained weights are deployed to the TRON system and further optimized in situ for another 25 epochs.  The NAS search space allows interlayer connections while enforcing a strictly feedforward computational graph, i.e., data flow is permitted only from earlier to later layers, with no connection from later to earlier layers. This ensures that the time-multiplexed execution is well-defined and can be carried out sequentially layer by layer.

For each task, we split the official training set into training and validation subsets, and only the training subset is used for parameter optimization. The validation subset is then used to evaluate the performance of each architecture candidate. During the whole in-situ NAS process, the test set remains untouched and is used only once for the final reported performance (accuracies in Fig.~\ref{fig:2}C, Fig.~\ref{fig:3}B, Tab.~\ref{tab:comparison}, and Fig.~\ref{fig:5}, Supplementary Section~8).

\subsection{Digital model of TRON unit}
Given that light propagation within a scattering medium is a linear and deterministic process, its transmission matrix can be physically modeled and characterized. We represent our system in a matrix form, wherein the output image can be formulated as follows:

\begin{equation}
I_{\text{out}} = |T_{\text{scattering}}E_{\text{in}}|^2 + \epsilon_{\text{noise}},
\label{eq:digitaltwin}
\end{equation}

where $E_{\text{in}}$ denotes the input DMD-modulated light field, $T_{\text{scattering}}$ represents the transmission matrix that characterizes the scattering medium, and $\epsilon_{\text{noise}}$ transmission matrix that characterizes the scattering medium, and additive measurement noise, respectively. We approximate the digital twin of the TRON unit by obtaining random input and output pairs and fitting them with the light propagation model described in Eq. \eqref{eq:digitaltwin}. Specifically, each $E_{\text{in}}$ is sampled independently from a Bernoulli distribution with $p=0.5$, and we sample $5\times10^4$ pairs. This fitting is performed once for a given task and the resulting digital twin is reused thereafter, rather than being redo per trial. The noise term is modeled using a multilayer perceptron with a hidden layer size of $2000$ and an architecture determined by another neural architecture search that optimizes the number of layers and activation functions. We train the digital twin by minimizing the mean squared error (MSE) loss using the Adam optimizer with a learning rate of $8\times 10^{-4}$, a batch size of $128$, and a StepLR scheduler with a step size of $20$ for $30$ epochs. These hyperparameters were selected based on a preliminary search for stable convergence and better fit quality. Once the digital twin is trained, we use it to provide approximate gradients during TRON optimization. The resulting gradient approximation and its use in training are described in the Supplementary Section~4.

\subsection{Training of deep TRON with backpropagation}
All instrumentation control and TRON training are implemented in Python (3.9.7). The TRON training is implemented using Pytorch's (1.10.0) automatic differentiation engine (autograd). Each unit in the TRON implements a controllable, parameterized forward transformation that maps an input pattern with parameters to an output intensity image measured by the camera. Because the DMD enforces binary modulation, the binarization step is non-differentiable. We therefore use a straight-through estimator (STE) with a $\text{HardTanh}$ surrogate in the backward pass to enable optimization through DMD binarization of both the input encoding and the trainable parameter encoding.  For other optical transformations, we employ a differentiable digital model that approximates each backward pass through a given model. The DMD area is divided into two sections to encode the input trainable parameters. During training and inference, the camera captures an intensity image per forward pass. For classification, the final captured image is partitioned into a number of regions equal to the number of classes. From each class region, we extract the intensity of the brightest speckle within that region. These class-wise intensities define the predicted class probabilities, and we optimize the network using loss $\text{CrossEntropy}$. The resulting loss is then sent through the digital twin to obtain approximate gradients with respect to the trainable parameters, which are updated using the $\text{Adam}$ optimizer and training hyperparameters specified by the in-situ NAS procedure.

\subsection{3D MedMNIST and RNA sequence classification task}
Description of the MedMNIST dataset: MedMNIST\cite{yang2023medmnist} is a large-scale MNIST-like collection of standardized biomedical images. Within this collection, six datasets are 3D and have been pre-processed into 28 $\times$ 28 $\times$ 28 (3D) dimensions with corresponding classification labels, without requiring prior background knowledge. Here, we evaluate TRON on the 3D MedMNIST organ classification tasks. For this task, the 3D images are transformed into 2D arrays and sequentially loaded onto TRON for training.

Description of RNA sequence data: The RNA sequence data are collected from a leukemia dataset in the Curated Microarray Database (CuMiDa) \cite{feltes2019cumida}, which has been carefully curated from more than 30,000 studies available in the Gene Expression Omnibus database. The data consists of seven distinct classes and encompasses a gene length of 222,284.

For both datasets and tasks, the data is binarized as it is loaded into the DMD. Once the neural architecture search is finished, we determine the best-performing architecture and then apply time-multiplexing to the data stream to enable inference. During inference, we carry out only the forward pass in situ and derive the classification outputs from the camera readout.

\subsection{Operation counting and throughput definitions}
We report forward-pass operation counts to enable a quantitative comparison with digital baselines. For a matrix-vector product between a matrix $W \in \mathbb{R}^{m\times n}$ and a vector $x \in \mathbb{R}^{n}$, we adopt the conventional count of $2mn$ scalar operations (one multiplication and one addition per entry of $W$). For TRON, these counts correspond to the effective dimensionality of the optical forward passes (Table~\ref{tab:comparison}) and do not include any electronic I/O, device control, or training-time gradient reconstruction. The raw optical projection rate is determined by the DMD refresh rate and camera frame rate, whereas the effective system throughput additionally depends on time-multiplexing (the number of optical units used to construct a given architecture) and task-specific utilization (the fraction of the DMD and camera region that is actively used by the task). We also note that, because the DMD and camera operate with 1-bit and 8-bit quantization, respectively, the optical operation counts should be interpreted as \emph{MAC-equivalent} figures rather than literal floating-point operations per second (FLOPS) as in the digital domain, because the dominant optical transformation is implemented by a large but fixed scattering matrix. The values reported in Table~\ref{tab:comparison} therefore provide an indicative arithmetic comparison under this convention. 

\section*{Acknowledgments}
F.X., Z.W., J.H., and S.G. acknowledge support from the Swiss National Science Foundation (SNF) project LION (216600). Z.W., F.X. and S.G. acknowledge  support from the LightOn Cloud for Research program, providing online and local access to LightOn’s OPU technology. Z.W. thanks the PSL Research University for their support. P.L.M. gratefully acknowledges support from a David and Lucile Packard Foundation Fellowship.

\section*{Author contributions:}
F.X. initiated the project. Z.W. and F.X. implemented the design with input from L.W., T.O., M.S., J.H., and P.M. F.X. and Z.W. prepared the manuscript. Z.W. and F.X. prepared the supplementary information. All authors read the article carefully and critically. S.G. and F.X. supervised the project.


\section*{Code and data availability:}
All data are available in the manuscript, the supplementary materials, or deposited online~\cite{github_repo}.

\bibliographystyle{npjqi}
\bibliography{references}

\end{document}